\documentclass[showpacs,preprintnumbers,aps,nofootinbib,floatfix,twocolumn]{revtex4}

\usepackage{epsfig}
\usepackage{citesort}

\begin{document}

\preprint{\begin{tabular}{l}
hep-ph/0402273, \\
KIAS-P04015
\end{tabular}}

\title{The effects of enhanced $Z$ penguins on lepton polarizations in
$B \to X_s \ell^+ \ell^-$}

  \author{S. Rai Choudhury}
     \email{src@physics.du.ac.in}
     \affiliation{Department of Physics \& Astrophysics \\
        University of Delhi, Delhi - 110 007, India}
  \author{A. S. Cornell}
   \email{alanc@kias.re.kr}
   \affiliation{Korea Institute of Advanced Study, Cheongryangri 2-dong, \\
      Dongdaemun-gu, Seoul 130-722, Republic of Korea}
   \author{Naveen Gaur}
     \email{naveen@physics.du.ac.in}
     \affiliation{Department of Physics \& Astrophysics \\
        University of Delhi, Delhi - 110 007, India}

\date{February $25^{th}$, 2004}


\begin{abstract}
The sensitivity of the $B \to \pi K$ mode to electro-weak penguins
and the recent experimental data for the $B \to \pi \pi, \pi K$
modes has given rise to what is known as the ``$B \to \pi K$
puzzle''. Recently it has been observed that this {\sl puzzle} can
be resolved by considering the new physics which can enter via
$Z^0$ penguins. In this note we analyze the effect of this
enhanced $Z^0$ penguins on the lepton polarization asymmetries of
$b \to s \ell^+ \ell^-$.
\end{abstract}

\maketitle

\par Recent observations of the $B$-meson decay into two pseudo-scalar
mesons have presented some significant deviations from the
currently available theoretical predictions pointed out in
\cite{Buras:2000gc} and recently reemphasized in
\cite{Yoshikawa:2003hb,Beneke:2003zv,Gronau:2003kj,Buras:2003dj}.
The decay into $\pi\pi$ channels can be reasonably well described
with the theoretical framework of the effective Hamiltonian
\cite{Buras:2003dj}, although the naive factorization used in this
context fails to describe the process and calculations involving
non-factorizable contributions which had to be included to yield
results in general agreement with data. Extensions of these
results using $SU(3)$ symmetries for decays into $\pi K$ channels,
however, show considerable disagreement with experimental values.
It has also been shown \cite{Buras:2003dj} that the results of
decays into $\pi K$ channels can be understood on the basis of an
enhanced $Z^0$ penguin diagram together with a large phase. This
may be the first indication of physics beyond the Standard Model
(SM) since the phase of the accepted values of the CKM matrix
cannot reproduce such a large phase in the $Z^0$ penguin diagram.

The possibility of strongly enhanced Z-penguin contributions was
carried in relation to $K \to \pi \nu \bar{\nu}$ and $K \to \pi
\ell \bar{\ell}$ decays for the first time in
\cite{Colangelo:1998pm} and in \cite{Buras:1998ed} where
constraints on these contributions imposed by the CP-violating
ratio $(\epsilon'/\epsilon)$ was investigated in a general class
of SUSY models in \cite{Buras:1999da}. The possibility of
non-standard $Z$ couplings in the context of $b \to s \bar{\ell}
\ell$ transitions was studied in \cite{Buchalla:2000sk}. However,
the authors of \cite{Buras:2003dj} were the first to relate
possible enhancement of the Z-penguin to the non-leptonic decay
modes involved in the``$B \to \pi K$'' puzzle and were able to
obtain definitive phenomenological values of the magnitude and
phase of $Z$-penguins consistent with the $B \to \pi \pi, \pi K$
data. The estimates of the magnitude and phase of the $Z^0$
penguin required to fit the $\pi\pi$ and $\pi K$ data, which have
been made in \cite{Buras:2003dj}, are purely on a phenomenological
basis. On the theoretical side such an enhancement of the penguins
can be accommodated within the supersymmetric extensions of the
SM, the Minimal Supersymmetric Standard Model (MSSM) in
particular. The flavour rotation of the squarks is different in
such theories from the corresponding flavour rotation of the
quarks, and this mismatch becomes the source of an additional
phase in flavour changing amplitudes. Any attempts to fit the
evaluated value of the $Z^0$ penguins with theory, however, is
hopeless, since the parameters involved in estimating the
resultant phase, which are essentially the off-diagonal elements
of the squark mass matrix, are not known \cite{Lunghi:1999uk}.
Irrespective of this the occurrence of a phase (other than the CKM
phase) in the $Z^0$ penguin is a signal for a new source of
CP-violation, and has wider implications.

\par The basic vertex involved in the analysis of \cite{Buras:2003dj}
is the $bsZ$ vertex. This vertex now having a phase beyond the CKM
one, will result in CP-violation in semi-leptonic decays of the
$B$-meson, such as $B \to X_s \ell^+ \ell^-$. As is well known,
due to the smallness of the coupling between $b$ and $u$, the $b
\to s \ell^+ \ell^-$ amplitude effectively has an overall CKM
phase. Therefore both in the SM and in supersymmetric extensions,
with only the CKM phase, information regarding the phase cannot be
extracted from any results involving this transition. As such we
shall point out in this brief note that with an effectively
complex $b s Z^0$ vertex the situation changes. In particular, we
show that possible measurements of polarization asymmetries in the
leptons, from the process $B \to X_s \ell^+ \ell^-$, will provide
a testing ground for the confirmation of the new phase and a
measurement of the magnitude of the $bsZ^0$ vertex, as marked out
in \cite{Buras:2003dj}.

\par The effective Hamiltonian for the $b \to s$ transition can be
written as
\begin{eqnarray}
{\cal H}_{eff}  &=&  \frac{\alpha G_F}{\sqrt{2} \pi}
V_{tb}V_{ts}^*
\left\{ C_9^{eff} \left( \bar{s} \gamma_{\mu} P_L b \right) \bar{\ell}
\gamma^{\mu} \ell  + C_{10} \left( \bar{s} \gamma_{\mu} P_L b \right)
 \right.  \nonumber  \\
&& \left. \times (\bar{\ell} \gamma^{\mu} \gamma^5 \ell)
- 2 C_7^{eff} \bar{s} i \sigma_{\mu \nu} \frac{q^{\nu}}{q^2}
m_b P_R b (\bar{\ell} \gamma^{\mu} \ell) \right\} . \nonumber \\
\label{eq:1}
\end{eqnarray}
where $q$ is the momentum transferred to the lepton pair, given as
$q = p_- + p_+$ (where $p_-$ and $p_+$ are the momenta of leptons
$\ell^-$ and $\ell^+$ respectively), $V_{tb} V_{ts}^*$ are the CKM
factors and $P_{L,R} = (1 \mp \gamma_5)/2$. The Wilson
coefficients $C_i$ are evaluated at the electroweak scale and then
evolved to the renormalization scale $\mu$. For our analysis with
the SM we choose a value for $C_7^{eff}$ and $C_{10}$ as:
$$ C_7^{eff} = -0.315 ~,~ C_{10} = - 4.642 .  $$
The coefficient $C_9^{eff}$ is complex within the SM model and is
a function of $\hat{s}$ ($= q^2/m_b^2$) in next-to-leading order,
where its value is given in \cite{Grinstein:1989me,Long-Distance}:
\begin{equation}
C_9^{eff} = C_9(\mu) + Y(\mu,\hat{s})
\end{equation}
where $Y(\mu,\hat{s})$ has the one-loop contributions of the four
quark operators, as given in \cite{Grinstein:1989me}. $C_9^{eff}$
also has a contribution from long distance effects associated with
the real $c\bar{c}$ resonances, where these are taken care of by
using the prescription given in \cite{Long-Distance,
Kruger:1996cv}.

\par From the expression of the matrix element given in
eqn.(\ref{eq:1}) we calculate the dilepton invariant mass distribution
as:
\begin{equation}
\frac{d \Gamma}{d\hat{s}} = \frac{G_F m_b^5}{192 \pi^3}
\frac{\alpha^2}{4 \pi^2} |V_{tb} V_{ts}^*|^2 (1 - \hat{s})^2 \sqrt{1 -
\frac{4 \hat{m_\ell}^2}{\hat{s}}} \bigtriangleup
\label{eq:2}
\end{equation}
where
\begin{eqnarray}
\bigtriangleup &=&
4  \frac{(2 + \hat{s})}{\hat{s}} \left(1 + \frac{2
\hat{m_\ell}^2}{\hat{s}}\right)
 |C_7^{eff}|^2 + (1 + 2 \hat{s})     \nonumber  \\
  && \left(1 + \frac{2 \hat{m_\ell}^2}{\hat{s}}\right) |C_9^{eff}|^2
+ \left(1 - 8 \hat{m_\ell}^2 + 2 \hat{s} + \frac{2
\hat{m_\ell}^2}{\hat{s}}\right)
\nonumber \\
&& \times |C_{10}|^2  + 12 \left(1 + \frac{2
\hat{m_\ell}^2}{\hat{s}}\right)
 Re(C_9^{eff *} C_7^{eff}) .
\label{eq:3}
\end{eqnarray}
\par To define the polarized branching ratio and then the polarization
asymmetries we will use the convention followed in earlier
references, such as \cite{Kruger:1996cv,RaiChoudhury:2002hf}. For
the evaluation of the polarized decay rates we introduce spin
projection operators, defined as $N = (1 + \gamma_5 {\not S}_x)/2$
for $\ell^-$ and $M = (1 + \gamma_5 {\not W}_x)/2$ for $\ell^+$,
where $x = L, N, T$ correspond to the longitudinal, normal and
transverse polarizations respectively. The orthogonal unit vectors
$S_x$ for $\ell^-$ and $W_x$ for $\ell^+$ in the rest frames of
respective leptons are defined as:
\begin{eqnarray}
S^\mu_L &\equiv& (0, {\bf e_L}) ~=~ \left(0,
\frac{{\bf p_-}}{|{\bf p_-}|} \right)            \nonumber     \\
S^\mu_N &\equiv& (0, {\bf e_N}) ~=~ \left(0, \frac{{\bf p_s} \times
{\bf p_-}}{|{\bf p_s} \times {\bf p_-}|}\right)    \nonumber      \\
S^\mu_T &\equiv& (0, {\bf e_T}) ~=~ \left(0, {\bf e_N} \times
\bf{e_L}\right)         \label{eq:4}             \\
W^\mu_L &\equiv& (0, {\bf w_L}) ~=~ \left(0, \frac{{\bf p_+}}{|{\bf
p_+}|}\right)             \nonumber                \\
W^\mu_N &\equiv& (0, \bf{w_N}) ~=~ \left(0, \frac{{\bf p_s} \times
                {\bf p_+}}{|{\bf p_s} \times {\bf p_+} |} \right)
                 \nonumber                \\
W^\mu_T &\equiv& (0, {\bf w_t}) ~=~ (0, {\bf w_N} \times {\bf w_L})
                 \label{eq:5}
\end{eqnarray}
where ${\bf p_-}$, ${\bf p_+}$ and ${\bf p_s}$ are respectively
the three momenta of $\ell^-$, $\ell^+$ and the $s$-quark in the
dileptonic c.m. frame. Note that the above polarization vectors
are defined in the rest frames of the leptons. We now boost these
to the dileptonic c.m. frame. Only longitudinal vectors which lie
along the boost will be boosted becoming:
\begin{eqnarray}
S^\mu_L &=& \left( \frac{|{\bf p_-}|}{m_\ell}, \frac{E_\ell {\bf
          p_-}}{m_\ell |{\bf p_-}|}   \right)      \nonumber      \\
W^\mu_L &=& \left( \frac{|{\bf p_-}|}{m_\ell}, - \frac{E_\ell {\bf
  p_-}}{m_\ell |{\bf p_-}|} \right) ,
\label{eq:6}
\end{eqnarray}
where $E_\ell$ is the energy of either lepton (where both leptons have
the same energy in the dileptonic c.m. frame).

\par The polarization asymmetries for $\ell^-$ are defined by the
equation
\begin{eqnarray}
{\cal P}_x^-(\hat{s}) & = & \frac{ d\Gamma ( \mathbf{n = e_x}
)/d\hat{s} -
d\Gamma ( \mathbf{n = -e_x} )/d\hat{s}}{d\Gamma ( \mathbf{n = e_x}
)/d\hat{s} + d\Gamma ( \mathbf{n = -e_x} )/d\hat{s}} .
\label{eq:7}
\end{eqnarray}
\begin{figure}
\includegraphics[width=3in]{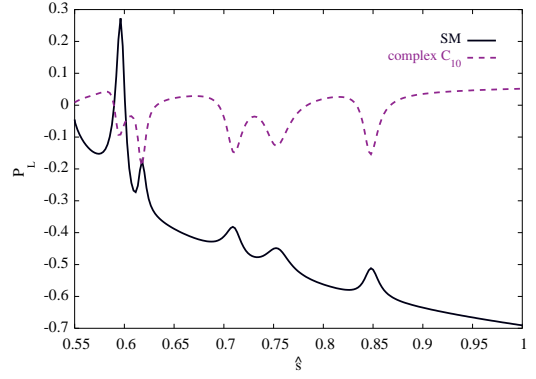}
\caption{Longitudinal polarization asymmetry (${\cal P}_L^-$) with
dilepton invariant mass (${\hat s}$)} \label{fig1}
\end{figure}

\begin{eqnarray}
{\cal P}_L^- & = & \frac{2}{\bigtriangleup} \left( 6 Re[C_{10}^*
C_7^{eff}] + \left(1 + 2 \hat{s} \right) Re[C_{10}^* C_9^{eff}]
\right)          \nonumber \\
&& \times\sqrt{ 1- \frac{4 \hat{m}_{\ell}^2}{\hat{s}}} ,
     \label{eq:8}      \\
{\cal P}_N^- & = & \frac{3 \pi \hat{m}_\ell}{2 \sqrt{\hat{s}}
 \bigtriangleup} \left( Im[C_7^{eff} C_{10}^*]
+ Im[C_9^{eff} C_{10}^*] \hat{s} \right) \nonumber \\
&& \times \sqrt{1 - \frac{4 \hat{m}_\ell}{\hat{s}}}  ,
     \label{eq:9}       \\
{\cal P}_T^- & = &
\frac{3 \pi \hat{m}_\ell}{2 \sqrt{\hat{s}} \bigtriangleup}
\left( - \frac{4}{\hat{s}} |C_7^{eff}|^2 - \hat{s} |C_9^{eff}| - 4
Re[C_7^{eff} C_9^{eff}]
\right.          \nonumber \\
&& \left. + 2 Re[C_7^{eff} C_{10}] + Re[C_9^{eff}
C_{10}] \right)
\label{eq:10}
\end{eqnarray}
The corresponding asymmetries for $\ell^+$ have expressions which
are identical to those above (apart from an overall negative sign
for ${\cal P}_L$ and ${\cal P}_N$), except for ${\cal P}_T$ where
the sign of the last two terms is changed.

\par In order to fit the data for $B \to \pi K$ it was proposed that
the $Z^0$ penguin has a large phase \cite{Buras:2003dj}. This
fitting was modelled by Buras {\em et al.} which effectively makes
the Wilson coefficient $C_{10}$ a complex valued:
\begin{equation}
C_{10}  =  - (2.2/\sin^2\theta_w) e^{ i \phi_{10}} ~~,~~ \phi_{10}
= \left(\frac{103}{180} \pi \right) , \label{eq:11}
\end{equation}
which not only has a large phase but also a magnitude more then twice
the SM expectation.

\par
\begin{figure}\includegraphics[width=3in]{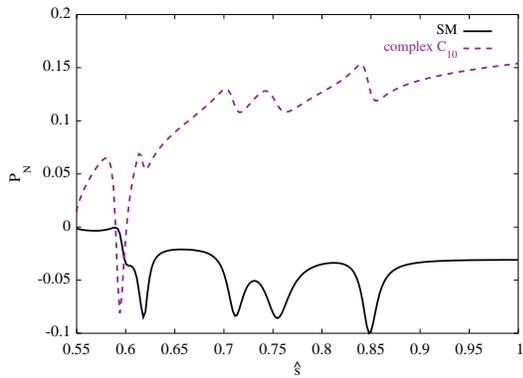}
\caption{Normal polarization asymmetry (${\cal P}_N^-$) with
dilepton invariant mass (${\hat s}$)} \label{fig2}
\end{figure}

\begin{figure}
\includegraphics[width=3in]{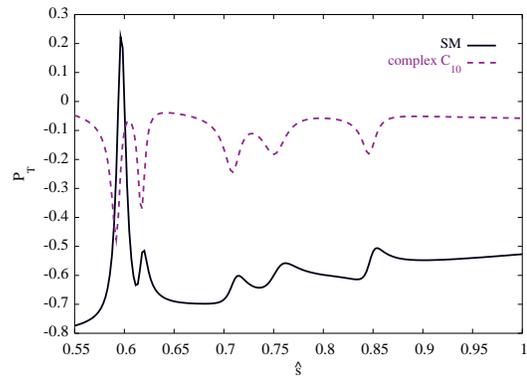}
\caption{Transverse polarization asymmetry (${\cal P}_T^-$) with
dilepton invariant mass (${\hat s}$)} \label{fig3}
\end{figure}

 In figures (\ref{fig1})-(\ref{fig3}) we show our results for the
various polarization asymmetries both within the SM and for the
enhanced value of the $Z^0$ penguin, as modelled in
eqn.(\ref{eq:11}). As can be seen the results dramatically change
with the new value of the coefficient $C_{10}$, as compared with
the results obtained in the SM and the MSSM. Measurement of these
polarization asymmetries would thus provide a direct test of the
validity of the model in \cite{Buras:2003dj} for enhanced and
complex values of the $bsZ$ vertex.  In table \ref{table:1} we
have given the averaged values of these asymmetries. The averaging
procedure which we have adopted is:\footnote{in the averaging we
have integrated the observables over $\hat{s}$ in the region which
is after the first charmonium resonance in the $b \to s \tau^+
\tau^-$ process.}
\begin{equation}
\langle {\cal P}_x \rangle = \frac{\displaystyle{\int_{(3.646 +
0.02)^2/m_b^2}^{1} {\cal P}_x \times \frac{d \Gamma}{d
\hat{s}} d \hat{s}}}{ \displaystyle{\int_{(3.646 +
0.02)^2/m_b^2}^{1}} \frac{d \Gamma}{d
\hat{s}} d \hat{s}}
\label{eq:12}
\end{equation}
As can see from the graphs in figures (\ref{fig1})-(\ref{fig3}) a
complex value of the $bsZ$ vertex gives the longitudinal and
transverse polarizations a decreased value as compared to their SM
values, however, the normal asymmetry shows a substantial increase
from its respective SM value. This can also be seen from the
averaged values of these asymmetries given in Table \ref{table:1}.
${\cal P}_N$ not only changes its sign as compared to its SM value
but also its magnitude increases by more than 100\%.

It is clear that future measurements of the enhanced normal
polarization asymmetry would be the more suitable testing ground for
the validity of a complex $bsZ$ vertex.
\begin{table}[h]
\begin{ruledtabular}
\begin{tabular}{l | c c  c  c }
Model &  BR $\times 10^7$ & ${\cal P}^-_L$ & ${\cal P}^-_N$
& ${\cal P}^-_T$  \\ \hline
SM & 2.39 & - 0.37  &   - 0.04  &  - 0.58    \\
enhanced $bsZ$ & 6.12 & -0.04  &   0.1  & - 0.13    \\
\end{tabular}
\caption{Predictions of the observables where BR is the branching
ratio of $B \to X_s \tau^+ \tau^-$} \label{table:1}
\end{ruledtabular}
\end{table}
\par We would like to make a few further observations. The
polarization asymmetries for $\ell^-$ in the CP conjugated process
$\bar{b} \to \bar{s} \ell^- \ell^+$ can be obtained from
eqns.(\ref{eq:8})-(\ref{eq:10}) by conjugating all the weak phases
whilst at the same time retaining the strong phases contained in
$C_9^{eff}$. If we call the polarization asymmetry of $\ell^-$ in
the conjugate process to be $\bar{{\cal P}}$ then in the SM we can
derive several relations, such as ${\cal P}_L = \bar{{\cal P}}_L$
and ${\cal P}_N = {\bar {\cal P}}_N$; these types of relations are
definitely violated if we have a large phase in $C_{10}$, which,
as in eqn.(\ref{eq:11}), makes it dominantly imaginary.

\par Atwood and Hiller \cite{Atwood:2003tg} have also recently
obtained an effectively enhanced $bsZ$ vertex. Their vertex
includes a right handed coupling. The non-zero phase obtained in
\cite{Buras:2003dj} is however quite rigid and a similar fit to
the available data, where a right handed vertex also is included,
has not been obtained.

\par For the decay $B \rightarrow X_d l^+ l^- $ the SM provides a
relative phase between the various contributing terms, and a
dominantly imaginary value of $C_{10}$ would cause substantial
interference between the CKM phases and the new $bsZ$ phase. Our
estimates of the polarization can easily be generalized to this
case, however, this decay is expected to be much weaker than the
one considered here and we do not present the results for this
mode in this note.

\par Finally, the recently published data on CP asymmetries in
the decay $B \rightarrow \phi K_s$
\cite{Abe:2003yt,Browder:2003ii} which has attracted a lot of
theoretical attention \cite{Kagan:1997sg}, in particular that of
Deshpande and Ghosh \cite{Deshpande:2003nx}, where they have
considered a model with extra down quarks and complex couplings,
obtaining bounds on the effective complex coupling parameters of
the $Z$-penguin graphs in the light of the data on $B\rightarrow
\phi K_s$ \cite{Abe:2003yt,Browder:2003ii}. In the type of
analysis that forms the basis of our polarization calculation,
Buras {\em et al.}\cite{Buras:2003dj} have considered in detail
the status of the effective penguin parameters obtained by them in
relation to the data on CP asymmetries in $B\rightarrow \phi K_s$.
The net result is that from the parameters obtained by them and
used by us, a value of $\sin 2\beta$ in the decay $ B\rightarrow
\phi K_s$ of the order +1, may well be possible. The experimental
data has large uncertainties with a value for this parameter given
as \cite{Abe:2003yt,Browder:2003ii}
$$ {\rm BaBar :}  + 0.45 \pm 0.43 \pm 0.07  $$
and
$$ {\rm Belle} : -0.96 \pm 0.50^{+0.09}_{-0.11} .$$
There is thus considerable disagreement between the average value
of the results of these two groups and the theoretical result
quoted above. However as observed in \cite{Buras:2003dj} the error
bars are too large for any definitive conclusion to be made. The
data of the two groups also has large differences. There are also
theoretical possibilities specific to the transition $b
\rightarrow ss\bar{s}$, such as a Higgs meditated amplitude in
SUSY models with large $tan\beta$ (considered by Kane {\em et
al.}\cite{Kagan:1997sg}) making it difficult to assess
definitively the disagreement between the predicted and the
experimental result for $B\rightarrow \phi K_s$. The scenario will
become clearer as more accurate experimental numbers become
available. Our results on the possibilities of polarization
results to study CP violation provide another parameter for
comparison when such results become available.


\acknowledgments{ SRC and NG's work was supported by Department of
Science \& Technology (DST), India, under project No.
SP/S2/K-20/99. ASC would also like to thank IRC (IUCAA reference
center), Delhi University and DST project no. SP/S2/K-20/99 for
providing local hospitality during his visit to Delhi, where this
work was initiated.}


\begin{thebibliography}{99}


\bibitem{Buras:2000gc}
 A.~J.~Buras and R.~Fleischer,
  Eur.\ Phys.\ J.\ C {\bf 16}, 97 (2000)
  [arXiv:hep-ph/0003323].

\bibitem{Yoshikawa:2003hb}
  T.~Yoshikawa,
  Phys.\ Rev.\ D {\bf 68}, 054023 (2003)
  [arXiv:hep-ph/0306147].

\bibitem{Gronau:2003kj}
  M.~Gronau and J.~L.~Rosner,
  Phys.\ Lett.\ B {\bf 572}, 43 (2003)
  [arXiv:hep-ph/0307095].

\bibitem{Beneke:2003zv}
  M.~Beneke and M.~Neubert,
  Nucl.\ Phys.\ B {\bf 675}, 333 (2003)
  [arXiv:hep-ph/0308039].

\bibitem{Buras:2003dj}
  A.~J.~Buras, R.~Fleischer, S.~Recksiegel and F.~Schwab,
  to appear in {\sl Phys. Rev. Lett.}
  [arXiv:hep-ph/0312259] .

  A.~J.~Buras, R.~Fleischer, S.~Recksiegel and F.~Schwab,
  arXiv:hep-ph/0402112.

  A.~J.~Buras, R.~Fleischer, S.~Recksiegel and F.~Schwab,
  Eur.\ Phys.\ J.\ C {\bf 32}, 45 (2003)
  [arXiv:hep-ph/0309012].

\bibitem{Colangelo:1998pm}
  G.~Colangelo and G.~Isidori,
  JHEP {\bf 9809}, 009 (1998)
  [arXiv:hep-ph/9808487];

\bibitem{Buras:1998ed}
A.~J.~Buras and L.~Silvestrini,
 Nucl.\ Phys.\ B {\bf 546}, 299 (1999)
 [arXiv:hep-ph/9811471].

\bibitem{Buras:1999da}
  A.~J.~Buras, G.~Colangelo, G.~Isidori, A.~Romanino and L.~Silvestrini,
  Nucl.\ Phys.\ B {\bf 566}, 3 (2000)
  [arXiv:hep-ph/9908371].

\bibitem{Buchalla:2000sk}
  G.~Buchalla, G.~Hiller and G.~Isidori,
  Phys.\ Rev.\ D {\bf 63}, 014015 (2001)
  [arXiv:hep-ph/0006136] ;

  G.~Isidori,
  arXiv:hep-ph/0009024.

\bibitem{Lunghi:1999uk}
  E.~Lunghi, A.~Masiero, I.~Scimemi and L.~Silvestrini,
  Nucl.\ Phys.\ B {\bf 568}, 120 (2000)
  [arXiv:hep-ph/9906286].

\bibitem{Grinstein:1989me}
  B.~Grinstein, M.~J.~Savage and M.~B.~Wise,
  Nucl.\ Phys.\ B {\bf 319}, 271 (1989) ;
  A.~J.~Buras and M.~M\"{u}nz,
  Phys.\ Rev.\ D {\bf 52}, 186 (1995)
  [arXiv:hep-ph/9501281].

\bibitem{Long-Distance}
  A.~Ali, T.~Mannel and T.~Morozumi,
  Phys.\ Lett.\ B {\bf 273}, 505 (1991);
  C.~S.~Lim, T.~Morozumi and A.~I.~Sanda,
  Phys.\ Lett.\ B {\bf 218}, 343 (1989);
  N.~G.~Deshpande, J.~Trampetic and K.~Panose,
  Phys.\ Rev.\ D {\bf 39}, 1461 (1989);
  P.~J.~O'Donnell and H.~K.~Tung,
  Phys.\ Rev.\ D {\bf 43}, 2067 (1991) .

\bibitem{Kruger:1996cv}
  F.~Kr\"{u}ger and L.~M.~Sehgal,
  Phys.\ Lett.\ B {\bf 380}, 199 (1996),
  [arXiv:hep-ph/9603237] ;
  J.~L.~Hewett,
  Phys.\ Rev.\ D {\bf 53}, 4964 (1996),
  [arXiv:hep-ph/9506289].
  %
  S.~Rai Choudhury, A.~Gupta and N.~Gaur,
  Phys.\ Rev.\ D {\bf 60}, 115004 (1999)
  [arXiv:hep-ph/9902355].
  %
  S.~Fukae, C.~S.~Kim and T.~Yoshikawa,
  Phys.\ Rev.\ D {\bf 61}, 074015 (2000)
  [arXiv:hep-ph/9908229].

\bibitem{RaiChoudhury:2002hf}
  S.~Rai Choudhury, N.~Gaur and N.~Mahajan,
  Phys.\ Rev.\ D {\bf 66}, 054003 (2002)
  [arXiv:hep-ph/0203041].
  %
  S.~R.~Choudhury and N.~Gaur,
  arXiv:hep-ph/0205076.
  %
  N.~Gaur,
  arXiv:hep-ph/0305242.

\bibitem{Atwood:2003tg}
  D.~Atwood and G.~Hiller,
  arXiv:hep-ph/0307251.

\bibitem{Abe:2003yt}
  K.~Abe {\it et al.}  [Belle Collaboration],
  Phys.\ Rev.\ Lett.\  {\bf 91}, 261602 (2003)
  [arXiv:hep-ex/0308035].

\bibitem{Browder:2003ii}
  T.~E.~Browder,
  Int.\ J.\ Mod.\ Phys.\ A {\bf 19}, 965 (2004)
  [arXiv:hep-ex/0312024].

\bibitem{Kagan:1997sg}
  A.~Kagan,
  arXiv:hep-ph/9806266.
  %
  G.~Hiller,
  Phys.\ Rev.\ D {\bf 66}, 071502 (2002)
  [arXiv:hep-ph/0207356].
  %
  A.~Datta,
  Phys.\ Rev.\ D {\bf 66}, 071702 (2002)
  [arXiv:hep-ph/0208016].
  %
  M.~Ciuchini and L.~Silvestrini,
  Phys.\ Rev.\ Lett.\  {\bf 89}, 231802 (2002)
  [arXiv:hep-ph/0208087].
  %
  B.~Dutta, C.~S.~Kim and S.~Oh,
  %
  Phys.\ Rev.\ Lett.\  {\bf 90}, 011801 (2003)
  [arXiv:hep-ph/0208226].
  %
  S.~Khalil and E.~Kou,
  %
  Phys.\ Rev.\ D {\bf 67}, 055009 (2003)
  [arXiv:hep-ph/0212023].
  %
  C.~W.~Chiang and J.~L.~Rosner,
  Phys.\ Rev.\ D {\bf 68}, 014007 (2003)
  [arXiv:hep-ph/0302094].
  %
  A.~Kundu and T.~Mitra,
  %
  Phys.\ Rev.\ D {\bf 67}, 116005 (2003)
  [arXiv:hep-ph/0302123].
  %
  K.~Agashe and C.~D.~Carone,
  Phys.\ Rev.\ D {\bf 68}, 035017 (2003)
  [arXiv:hep-ph/0304229].
  %
  G.~L.~Kane, P.~Ko, H.~b.~Wang, C.~Kolda, J.~h.~Park and L.~T.~Wang,
  Phys.\ Rev.\ Lett.\  {\bf 90}, 141803 (2003)
  [arXiv:hep-ph/0304239].
  %
  D.~Chakraverty, E.~Gabrielli, K.~Huitu and S.~Khalil,
  Phys.\ Rev.\ D {\bf 68}, 095004 (2003)
  [arXiv:hep-ph/0306076].
  %
  J.~F.~Cheng, C.~S.~Huang and X.~h.~Wu,
  %
  Phys.\ Lett.\ B {\bf 585}, 287 (2004)
  [arXiv:hep-ph/0306086].
  %
  R.~Arnowitt, B.~Dutta and B.~Hu,
  Phys.\ Rev.\ D {\bf 68}, 075008 (2003)
  [arXiv:hep-ph/0307152].
  %
  C.~Dariescu, M.~A.~Dariescu, N.~G.~Deshpande and D.~K.~Ghosh,
  %
  arXiv:hep-ph/0308305.

\bibitem{Deshpande:2003nx}
  N.~G.~Deshpande and D.~K.~Ghosh,
  arXiv:hep-ph/0311332.

\end{thebibliography}
\end{document}